\documentclass[aip,jcp,a4paper,12pt,nofootinbib]{revtex4-1}

\usepackage[utf8]{inputenc}
\usepackage[T1]{fontenc}

\usepackage{amsmath}
\usepackage{amssymb,amsfonts,bm,bbm} 
\usepackage{hyperref}
\usepackage{graphicx,color}
\usepackage{units}
\usepackage{mathtools}

\newcommand{\myinclude}[2]{{\includegraphics{pics/ellipses_a_N10_Ex3}}}

\newcommand{\jl}{[\![}
\newcommand{\jr}{]\!]}
\renewcommand{\jmp}[1]{\jl#1\jr}

\newcommand{\Oi}{\Omega_i}
\newcommand{\Oj}{\Omega_j}

\newcommand{\Ozc}{\overline{\Omega}_0^{\mathsf c}}

\newcommand{\Oz}{\Omega_0}

\newcommand{\Gi}{\Gamma_i}
\newcommand{\Gj}{\Gamma_j}

\newcommand{\Gz}{\Gamma_0}

\newcommand{\IntM}{\{1,\ldots,M\}}

\newcommand{\real}{\mathbb R}

\newcommand{\grad}{\nabla}

\newcommand{\Ylm}{\mathcal Y_{\ell m}}

\newcommand{\Yzz}{\mathcal Y_{0 0}}
\newcommand{\Ylpmp}{\mathcal Y_{\ell ' m'}}

\makeatletter
\newcommand{\pushright}[1]{\ifmeasuring@#1\else\omit\hfill$\displaystyle#1$\fi\ignorespaces}
\newcommand{\pushleft}[1]{\ifmeasuring@#1\else\omit$\displaystyle#1$\hfill\fi\ignorespaces}
\makeatother

\newcommand{\SLO}{\mathcal S}

\newcommand{\lambdav}{\bm\lambda}

\newcommand{\Psiv}{\bm{\Psi}}

\usepackage{filecontents}

\begin{document}
\title{Theoretical analysis of screened many-body electrostatic interactions between charged polarizable particles}

\author{Eric B. Lindgren}
\affiliation{Aachen Institute for Advanced Study in Computational Engineering Science (AICES), RWTH Aachen University, Schinkelstr. 2, 52062 Aachen, Germany}
\affiliation{Center for Computational Engineering, Mathematics Department, RWTH Aachen University, Schinkelstr. 2, 52062 Aachen, Germany}
\author{Chaoyu Quan}
\affiliation{Sorbonne Universit\'es, UPMC Univ Paris 06, UMR 7598, Institut des Sciences du Calcul et des Donn\'ees, F-75005, Paris, France}
\author{Benjamin Stamm}
\affiliation{Center for Computational Engineering, Mathematics Department, RWTH Aachen University, Schinkelstr. 2, 52062 Aachen, Germany}

\begin{abstract}
\small
	
\begin{center}
	\textbf{Abstract}
\end{center}

This paper builds on two previous works, Lindgren et al. \textit{J. Comp. Phys.} 371, 712-731 (2018) \& Quan et al. arXiv:1807.05384 (2018), to devise a new method to solve the problem of calculating electrostatic interactions in a system composed by many dielectric particles, embedded in a homogeneous dielectric medium, which in turn can also be permeated by charge carriers. The system is defined by the charge, size, position and dielectric constant of each particle, as well as the dielectric constant and Debye length of the medium. The effects of taking into account the dielectric nature of the particles is explored in selected scenarios where the presence of electrolytes in the medium can significantly influence the total undergoing interactions. Description of the mutual interactions between all particles in the system as being truly of many-body nature reveals how such effects can effectively influence the magnitudes and even directions of the resulting forces, especially those acting on particles that have a null net charge. Particular attention is given to a situation that can be related to colloidal particles in an electrolyte solution, where it's shown that polarization effects alone can substantially raise or lower---depending on the dielectric contrast between the particles and the medium---the energy barrier that divides particle coagulation and flocculation regions, when an interplay between electrostatic and additional van der Waals forces is considered. Overall, the results suggest that for an accurate description of the type of system in question, it is essential to consider particle polarization if the separation between the interacting particles are comparable to or smaller than the Debye length of the medium.

\normalsize
\end{abstract}

\maketitle

\section*{Introduction}

Electrostatic interactions are of great relevance in many areas of science and technology. Examples range from the kinetic nonlability of colloids in suspension, rendered by the presence of like charges on their surfaces, \cite{Atkins,Israelachvili} on to the behaviour of charged grains in dusty plasmas, \cite{Mendis1994,LampeDustyPlasma} and through to the self-assembly of nanocrystals, influenced by opposite-charge interactions. \cite{Shevchenko,Kovalenko2015} In many systems of interest, the medium in which particles are embedded is often permeated by charge carries, such as protons, electrons or electrolytes, with the ability to screen electrostatic interactions, effectively provoking a faster decay of the electric potential generated by each charged particle of the system. Strictly speaking, even pure water is an electrolyte solution, since it contains 10$^{-7}$ molar (mol$\cdot$dm${^{-3}}$) of both hydronium (H$_3$O$^+$) and hydroxide (OH$^-$) ions, which translates to a Debye length of approximately 1 $\mu$m \cite{Israelachvili}. One way to account for the presence of charged species in the medium is to describe them explicitly which, in turn, can be in practice very computationally expensive. An alternative formulation consists of a continuous description of these charges. Accordingly, the electrostatic component of the interaction between $ n $ point-charges can be described within a continuous medium by the screened Coulomb (Yukawa) potential, \cite{Leimkuhler2016}

\begin{equation*}
	\Phi^{\textrm{Yukawa}} = \dfrac{K}{\varepsilon_\text{m}} \sum_{i=1}^{n} \sum_{j > i}^{n} \dfrac{q_i q_j}{r_{ij}} e^{-\kappa r_{ij}},
\end{equation*}

\noindent where $K$ is the Coulomb's constant, $\varepsilon_\text{m}$ is the dielectric constant (relative permitivitty) of the medium, $q_i$ and $q_j$ are the respective charges participating in a pairwise interaction, $r_{ij}$ is the center-to-center separation between these particles, and $\kappa^{-1}$ is the Debye length---a characteristic decay length that depends exclusively on the properties of the medium. The Yukawa potential consists of a limiting case of the well established DLVO theory, \cite{Hopkins2005} which further accounts for the contributions of van der Waals forces and the finite size of the interacting particles. Developed  by B. Derjaguin and L. Landau, \cite{DerjaguinLandau1941} and independently by E. Verwey and J.T.G. Overbeek, \cite{VerweyOverbeek1948} DLVO theory is still vastly employed today in the understanding of screened electrostatic interactions, particularly in colloidal chemistry. There are many situations, however, where DLVO theory fails to provide an appropriate description. Commonly described in the literature as ``non-DLVO" forces, other contributions to the overall interaction may be too important to be neglected in various scenarios. Examples of such contributions include forces of steric and of solvation nature. \cite{Israelachvili,Grasso2002}

Another contribution that is commonly overlooked in descriptions of screened electrostatic interactions is that of polarization---a direct response from a particle to an external electric field, which translates to an anisotropy of its surface charge. Recent studies have shown that the effect of particle polarization can drastically change the nature of an interaction, with extreme cases being notably the ones where particles with same sign of charge attract one another (event that is generally dependant on certain degrees of asymmetry in particles' size or charge, and a short inter-particle separation distance), and cases where particles with opposite sign of charge repel one another (which involves magnitudes of particles' charge that are below some critical limit and the presence of a sufficiently polarizable medium). \cite{Linse,Bichoutskaia,Stace2011,Lekner,Filippov,LindgrenTitan,LindgrenMedium} In systems that count with the presence of a medium that is polarizable ($\varepsilon_\text{m} > 1$) and charge carriers such as, for instance, aqueous electrolyte solutions, not only long-ranged Coulomb (monopole) forces between charged interacting particles decay more pronouncedly, but polarization forces, that already naturally decay faster than the former, are more effectively suppressed, as the latter rely on the ability of a particle to experience the electric potential generated by another, which suffers from screening in this situation. Ergo, the neglect of such polarization effects being justified as a reasonable approximation. There are situations, however, where particle polarization should be accounted for, as demonstrated in a recent paper by Derbenev et al. \cite{Derbenev} that shows that for microsized PMMA spherical particles in a low-dielectric solvent permeated by screening agents, polarization forces can be comparable to van der Waals' at short inter-particle separations, and the interaction significantly differs from that predicted by DLVO theory. A further contribution to the overall interaction, tied to particle polarization, is that provided by many-body effects. Such contribution is potentially important in systems where three or more particles interact with one another and is of the same nature of that involving the induction component of intermolecular forces, which is recognized as being strongly nonadditive, since the fields of the different interacting neighbouring molecules may reinforce or cancel out each other. \cite{Stone,Batista2015nonadditivity} As demonstrated elsewhere, \cite{LindgrenManyBody2018} many-body effects due to mutual polarization are rooted on the ability of each particle to become polarized when they experience an external electric field, as it occurs in a simple two-body interaction, but with the addition of being also heavily dependant on particles' relative position.

This work presents a new solution to the problem of calculating electrostatic interactions between charged polarizable particles in a many-body system, in the presence of a polarizable medium that can also be permeated by charge carriers. The model and numerical method, summarized in the following section and thoroughly described in Appendix A, are derived and generalized from two previous and related models, Lindgren et al. \cite{LindgrenManyBody2018} and Quan et al. \cite{quan2018domain}, thus combining their relevant capabilities in one package. Such capabilities are presented in terms of numerical examples in the context of colloidal chemistry, where it is sought to determine to what extent it is important to consider polarization and many-body effects in screened electrostatic interactions.

\section*{Model overview} 
\label{model_overview}

\begin{figure}[h]
	\centering
	\includegraphics[width=0.3\textwidth]{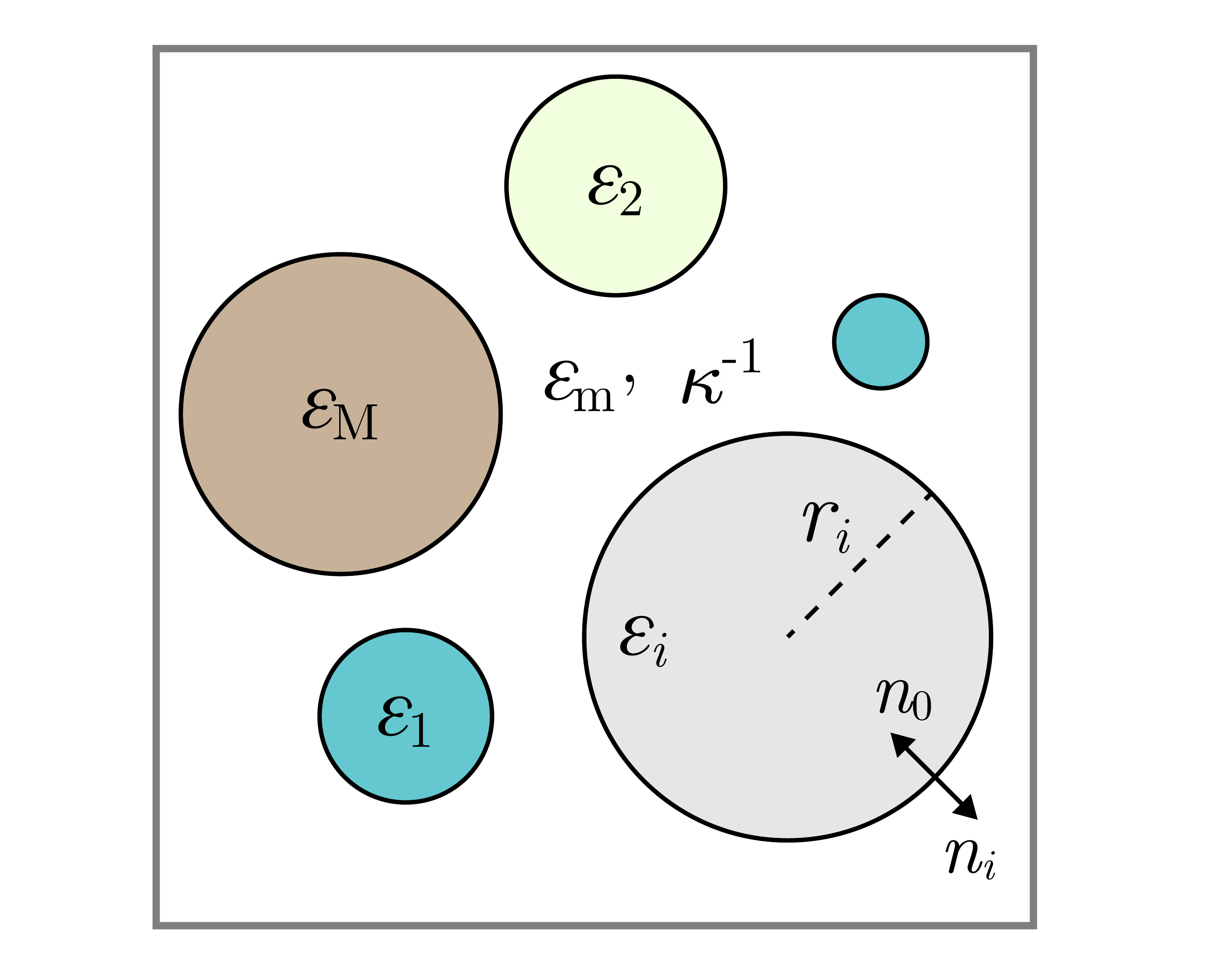}
	\caption{Graphical representation of the system.}
	\label{fig:Geom}
\end{figure}

The considered model can be described as a collection of $M$ non-overlapping spherical particles $\Omega_i$ ($i=1,\ldots,M$) in $\mathbb R^3$, where each $\Omega_i$ has radius $r_i$ and is centered at $x_i\in\mathbb R^3$, as illustrated in Figure \ref{fig:Geom}. The dielectric constants (relative permittivity) of each particle $\Omega_i$ and of the surrounding medium $\Oz$ are denoted by $\varepsilon_i$ and $\varepsilon_\text{m}$, respectively. The medium can be permeated by a continuum of charge carriers which define a particular value of Debye length ($\kappa^{-1}$), ranging from infinity, when such charge carriers are completely absent, to just a few nanometers, as in the case of concentrated electrolyte solutions. 

Now, denote the medium, i.e. the complement to the particles, by $\Oz=\real^3\backslash \cup_{i=1}^M \overline{\Omega}_i$ and its boundary by $\Gamma_0 = \partial \Oz$. 
Further, denote the particles' surfaces by $\Gamma_i=\partial\Omega_i$ for $i=1,\ldots,M$.
Then, we have the following relation
\begin{align*}
	\Ozc &= \Omega_1\cup \ldots \cup \Omega_M,\\
	\Gamma_0 &=\Gamma_1\cup \ldots \cup \Gamma_M.
\end{align*}

\noindent The piecewise dielectric constant, $\varepsilon$, can be written as a spatial function
\[
\varepsilon(x) 
= \varepsilon_\text{m} 
+ \sum_{i = 1}^M (\varepsilon_i-\varepsilon_\text{m})\mathbbm{1}_{\Omega_i}(x), \qquad x\in \real^3,
\]
where $\mathbbm{1}_{\Omega_i} $ denotes the characteristic function of $\Omega_i$. 
Each particle is assumed to carry a surface free charge $q_i$, which can also be represented by an isotropic surface density 
$$\sigma_{f,i} = q_i / (4 \pi r_i^2) \in\mathbb R.$$
Let $\sigma_f$ denote the global density as follows
\begin{equation}
	\sigma_f(x) = 
	\begin{cases} \sigma_{f,i} & \mbox{if } x\in \Gi,\\
		0 & \mbox{otherwise}.
	\end{cases}\label{eq:sigma_f}
\end{equation}

\noindent The electrostatic problem is modeled by a linearized Poisson-Boltzmann equation defined in the medium, $\Oz$, and a Laplace equation defined within the collection of particles, $\Ozc$, i.e.,
\begin{align}
	\label{eq:PDE_out}
	- \varepsilon_\text{m} \nabla \cdot \Phi(x) + \kappa^2(x) \Phi(x) = 0 & \quad \text{in} \; \Oz, 
	\\
	\label{eq:PDE_in}
	\Delta \Phi(x) = 0  & \quad \text{in} \; \Ozc,
\end{align}
\noindent with the following boundary conditions
\begin{align}
	\label{eq:BC1}
	\jmp{\Phi} &= 0 \qquad \quad \; \; \text{on } \Gamma_0, 
	\\
	\label{eq:BC2}
	\jmp{\varepsilon \grad \Phi } &= 4 \pi K \sigma_f \quad \text{on } 	\Gamma_0.
\end{align}
\noindent Here, $K$ is the Coulomb's constant, and $\jmp{\Phi}$ and $\jmp{\varepsilon \grad \Phi }$ are jumps defined by
\begin{align*}
	\jmp{\Phi}|_{\Gamma_i} (x) &= \Phi|_{\Omega_0} (x) n_0(x) + \Phi|_{\Omega_i} (x) n_i(x),\\
	\jmp{\varepsilon \grad \Phi}|_{\Gamma_i} (x) &= (\varepsilon\nabla \Phi)|_{\Omega_0} (x) \cdot n_0(x) + (\varepsilon\nabla \Phi)|_{\Omega_i} (x) \cdot n_i(x),
\end{align*} 
where $n_i(x)$ and $n_0(x)$ denote the unit normal vectors respectively pointing outward $\Omega_i$ and $\Omega_0$, for all $x\in\Gamma_i$ ($i=1,\ldots,M$).

The potential $\Phi$ can be uniquely represented in terms of an integral equation and a solution to this problem can be obtained by means of Galerkin discretization based on a truncated series of real spherical harmonics of maximum degree $ N $, as described in Appendix A, where it is also introduced how relevant physical properties of the system, namely the electrostatic energy of the system and the electrostatic force acting on each particle, are obtained. In all the following calculations the maximum degree is fixed to $ N = 15 $, a value that ensures numerical accuracy for the calculated cases.

\section*{Results and discussion}

\subsection*{Electric potential}

We first explore how the electric potential in an electrolyte solution is influenced by the ionic strength and the polarization effects. Consider two 50 nm-radius particles, with dielectric constants $\varepsilon_i=80/4=20$ or $\varepsilon_i=80\times4=320$ ($i=1,2$), surface charge densities $\sigma_{f,1} = -\sigma_{f,2} = -0.0025$ e$\cdot$nm$^{-2}$, separated by a center-to-center distance of 200 nm, and immersed in an aqueous electrolyte solution with $\varepsilon_\text{m} = 80$. The system is anti-symmetric, as the particles are identical apart from their signs of charge, and a 1:1 electrolyte provides equal amount of positive and negative ions. In this particular scenario, two screening effects are in operation, namely the one exerted by the dielectric medium (water), and the one provided by the charge carriers (electrolytes). Regarding the first, the polarizable medium is taken to be homogeneous and therefore would weaken the magnitude of the electric potential at any point in $\Oz$ by a factor that has the same value of its dielectric constant, i.e. 80, if the interaction was purely of Coulombic nature. 
As particle polarization is involved, such observation is virtually true only for points that are not at the vicinity of the particles. Indeed, in neighboring regions to charged polarizable particles, attenuation of the electric potential by a dielectric medium generally does not follow such simple observation that is valid for pure Coulombic interactions, as addressed in more details elsewhere \cite{LindgrenMedium}. With respect to the second screening effect, the cation and anion derived from the 1:1 electrolyte tend to be more concentrated around the respective particles of opposite charge, therefore the electric potential falls more pronouncedly closer to the charged surfaces $\Gamma_i$ ($i=1,2$). 

\begin{figure}[h]
	\centering
	\includegraphics[width=0.9\textwidth]{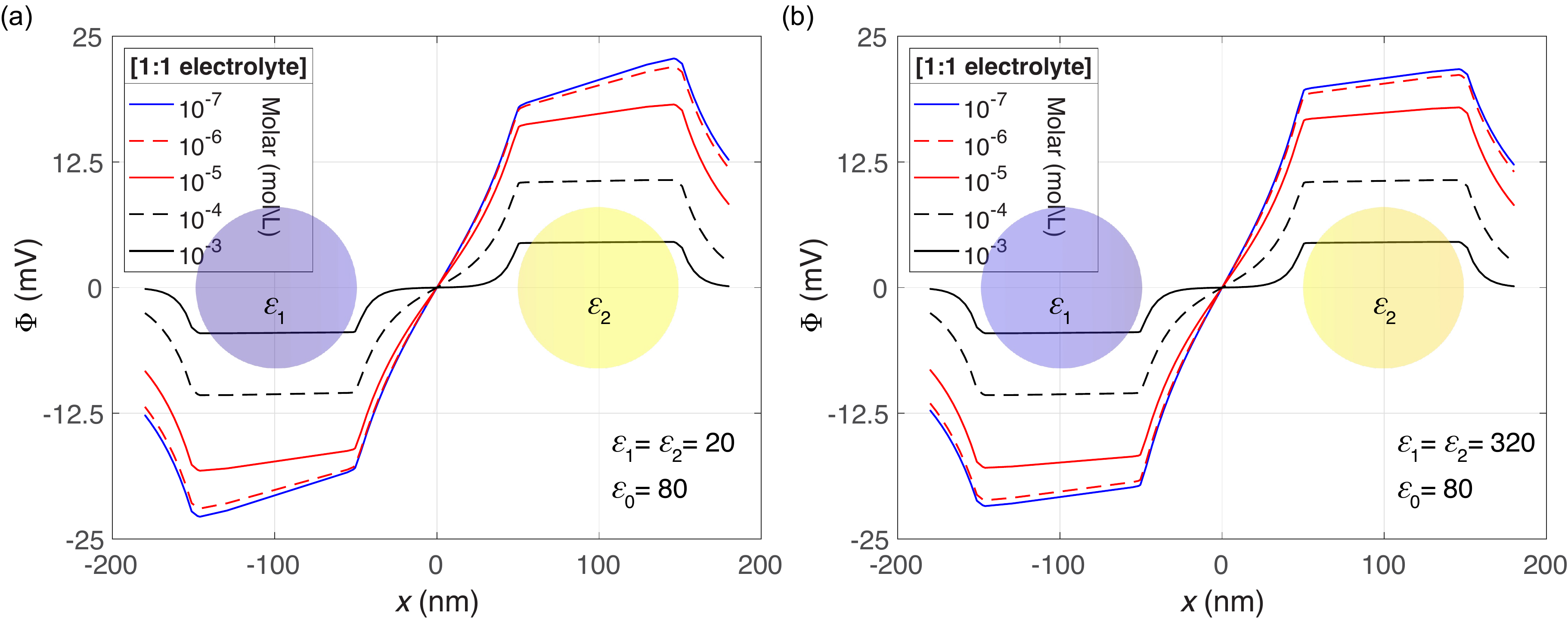}
	\caption{Plot of the electric potential $\Phi$ (mV) along the \textit{x}-axis that passes through the centers of two 50 nm-radius oppositely charged particles, where $\sigma_{f,1} = -\sigma_{f,2} = -0.0025$ e$\cdot$nm$^{-2}$ and $\varepsilon_1=\varepsilon_2=20$ (a) or $\varepsilon_1=\varepsilon_2=320$ (b), for various concentrations of a 1:1 electrolyte. The solvent is water, with $\varepsilon_\text{m} = 80$.}
	\label{fig:Potential2S}
\end{figure}

These effects are reflected in Figure \ref{fig:Potential2S}, which shows a plot of the electric potential calculated at points along the axis joining the two particles for different ionic strengths, the latter being determined by the molar concentration of a 1:1 electrolyte (e.g. NaCl), where the relation \cite{Israelachvili} $ \kappa^{-1} = 0.304/[\text{1:1 electrolyte}/\text{molar}]$ nm has been employed to extract the corresponding Debye length of the medium, considering the system is at 298K. When the solvent is pure water, [electrolyte] = $10^{-7}$ molar, the screening effect is at its minimum and particle polarization can take place in a considerable manner, resulting in anisotropy of the distribution of surface charge. 
As the medium are either more (Figure \ref{fig:Potential2S}a) or less (Figure \ref{fig:Potential2S}b) polarizable than the particles, these experience polarization that is strongly mediated by the former, with the creation of bound charges at the interface $\Gz = \Gamma_1 \cup \Gamma_2$, which turns each particle boundary effectively less charged on the side that is close to the other particle and more charged on the side faraway when $\varepsilon_1 = \varepsilon_2 < \varepsilon_\text{m}$, or the opposite when $\varepsilon_1 = \varepsilon_2 >\varepsilon_\text{m}$, while maintaining their given free charge constant in both cases. Consequently, the magnitude of the electric potential across each particle and towards the inter-particle region decreases more pronouncedly in the first case than in the second. 
As the solution becomes more concentrated, the electric potential is more effectively screened and progressively flattens out across both particles. 
In this situation, the given free charge on one particle contributes less effectively to the potential at the vicinity of the other particle, also influencing the extent in which the latter is polarized, for such effect depends on the ability of a particle to be in reach of a sufficiently strong electric field created by any external (to self) charge. As a consequence, the contribution of bound charges to the total electric potential is also weakened.

\begin{figure}[h]
	\centering
	\includegraphics[width=0.9\textwidth]{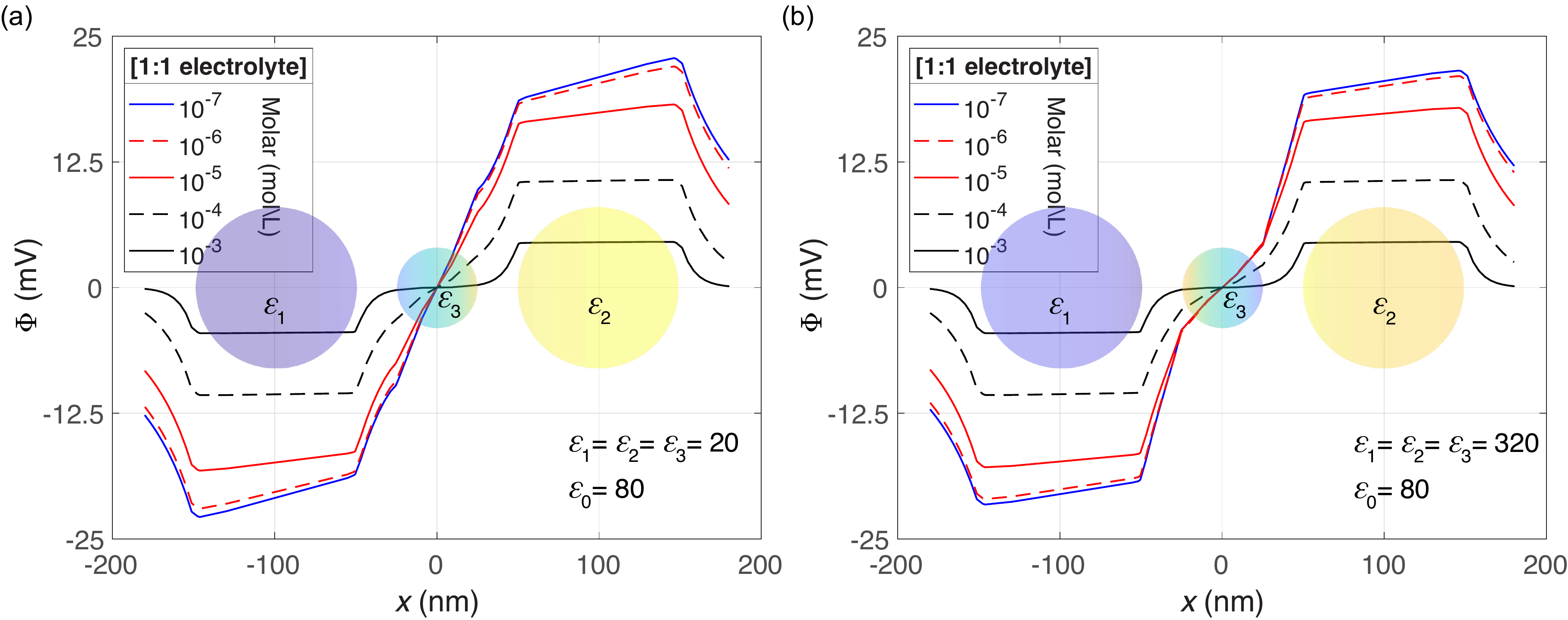}
	\caption{Plot of the electric potential $\Phi$ (mV) along the \textit{x}-axis that passes through the centers of two 50 nm-radius oppositely charged particles, interposed by a 25 nm-radius neutral particle, where $\sigma_{f,1} = -\sigma_{f,2} = -0.0025$ e$\cdot$nm$^{-2}$ and $\varepsilon_1=\varepsilon_2=\varepsilon_3=20$ (a) or $\varepsilon_1=\varepsilon_2=\varepsilon_3=320$ (b), for various concentrations of a 1:1 electrolyte. The solvent is water, with $\varepsilon_\text{m} = 80$. }
	\label{fig:Potential3S}
\end{figure}

Consider now adding a third, 25 nm-radius neutral particle to the system, positioned in between the two original charged species and having the same dielectric constant, i.e. $\varepsilon_1=\varepsilon_2=\varepsilon_3=20$ or $\varepsilon_1=\varepsilon_2=\varepsilon_3=320$. The presence of the neutral particle disrupts the electric potential of the original configuration, as shown in Figure \ref{fig:Potential3S}, and such effect is naturally more pronounced for lower electrolyte concentrations, for in these cases the magnitude of the potential is higher in the neighboring regions to the neutral particle, which is then more effectively polarized. 
When the neutral particle is more polarizable than the medium (Figure \ref{fig:Potential3S}b), it symmetrically acquires a partial negative bound charge on the side that faces the positive particle and a partial negative bound charge on the other side that faces the positive particle, as would be expected if the interaction took place in free space (vacuum). In this case, the presence of the neutral particle causes a faster decay of the potential magnitude in the original inter-particle region, as much alike as an insulator shields the electric potential in a parallel plate capacitor. 
However, when the neutral particle is less polarizable than the medium (Figure \ref{fig:Potential3S}a), the interface neutral particle--medium gets polarized in reverse, i.e. a partial positive bound charge now appears on the side that faces the positive particle and a partial negative bound charge appears on the other side that faces the negative particle.
This \textit{switch} effect has been addressed in more details elsewhere in the literature \cite{Israelachvili,LindgrenMedium}.

The above results suggest that the electric potential over some regions can be modulated by the parameters explored, namely the piecewise dielectric constant across the system and the ionic strength of the medium. The effect brought about by the latter is straightforward while the ones related to the former are more intricate. Starting then with the latter, it can be anticipated that an increase in electrolyte concentration will result in a stronger screening effect which, as a consequence, translates to a faster decay of the electric potential generated by the each charged particle of the system. Such effect also influences the extent in which each dielectric particle gets polarized, for it is related to the magnitude of the electric field experienced by the particle.

Concerning the piecewise dielectric constant across the system, for a homogeneous medium, $\varepsilon_\text{m}$ indicates the approximate magnitude in which the electric potential is attenuated in $ \Omega_0 $. 
For finite-size dielectric particles such indication is only an approximation since, particularly at points localized at the vicinity of the particles, there's an additional factor related to the medium and also to the particles, namely the contrast between $\varepsilon_i$ and $\varepsilon_\text{m}$ (for each $i=1,2,\dots,M$), that locally influences the electric potential by determining the extent and form in which particle polarization takes place. 
For any $\varepsilon_i = \varepsilon_\text{m}$, the lack of dielectric discontinuity across each $\Gi$ would ensure that the particles are not polarized and in such case, the system could be appropriately described by DLVO theory. However, the presence of a dielectric discontinuity across each $\Gi$ in cases where $\varepsilon_i < \varepsilon_\text{m}$ or $\varepsilon_i > \varepsilon_\text{m}$ indicates one of the two objects, medium or particle, is more polarizable than the other, which translates to the appearance of bound charge in such boundary, making it necessary to consider induced multipoles for an accurate description. 
As a general rule, the higher the contrast between $\varepsilon_i$ and $\varepsilon_\text{m}$, the more significant is the induction of bound charge at the respective surface $\Gi$ when particle $i$ is exposed to an external electric field. Further, it is recognized that higher values of dielectric constant ($k \gtrapprox 1000$) causes a dielectric particle to start responding to an external electric field in a similar fashion to that of metallic particles \cite{Filippov,LindgrenPerspective}, i.e. by asymptotically zeroing the field in its interior. Conductors are able to cancel the field in its interior, but such cancellation within a dielectric is incomplete, being related to the ability of a particle to be polarized. 

Therefore, considering as an example the geometry addressed in Figure \ref{fig:Potential2S}, the possible effects of polarization on the electric potential can be summarized as follows, taking as a comparative reference the case where the polarization is suppressed, i.e. $\varepsilon_1 = \varepsilon_2 = \varepsilon_\text{m}$. 
When $\varepsilon_1 = \varepsilon_2 < \varepsilon_\text{m}$, the potential has its magnitude decreased more pronouncedly across each $\Oi$ towards the region in between particles. In other words, the variation of the potential inside $ \Omega_i $ ($ i=1,2 $) is greater. When $\varepsilon_1 = \varepsilon_2 > \varepsilon_\text{m}$, the magnitude of the potential also decreases across each $\Oi$ towards the region in between particles, but in a less pronounced manner, and then becomes progressively flattened as $\varepsilon_1 = \varepsilon_2 \rightarrow \infty$.

\subsection*{Electrostatic force}

Calculations were also performed to investigate the influence which polarization and screening effects can have on the electrostatic force in a system composed of many dielectric particles. 
Consider three representatives of positive, negative and neutral particles, randomly displaced  in a aqueous electrolyte medium, having their positions restricted to the x-y plane (for a visual clarity of results). Each neutral particle has a radius of 75 nm, while the positive and negative particles each has a radius of 50 nm and a magnitude of surface charge density 0.01 e$\cdot$nm$^{-2}$. The electrostatic force acting on each particle is then explored in terms of their dielectric constants and the ionic strength of the medium.

\begin{figure}[h]
	\centering
	\includegraphics[width=0.72\textwidth]{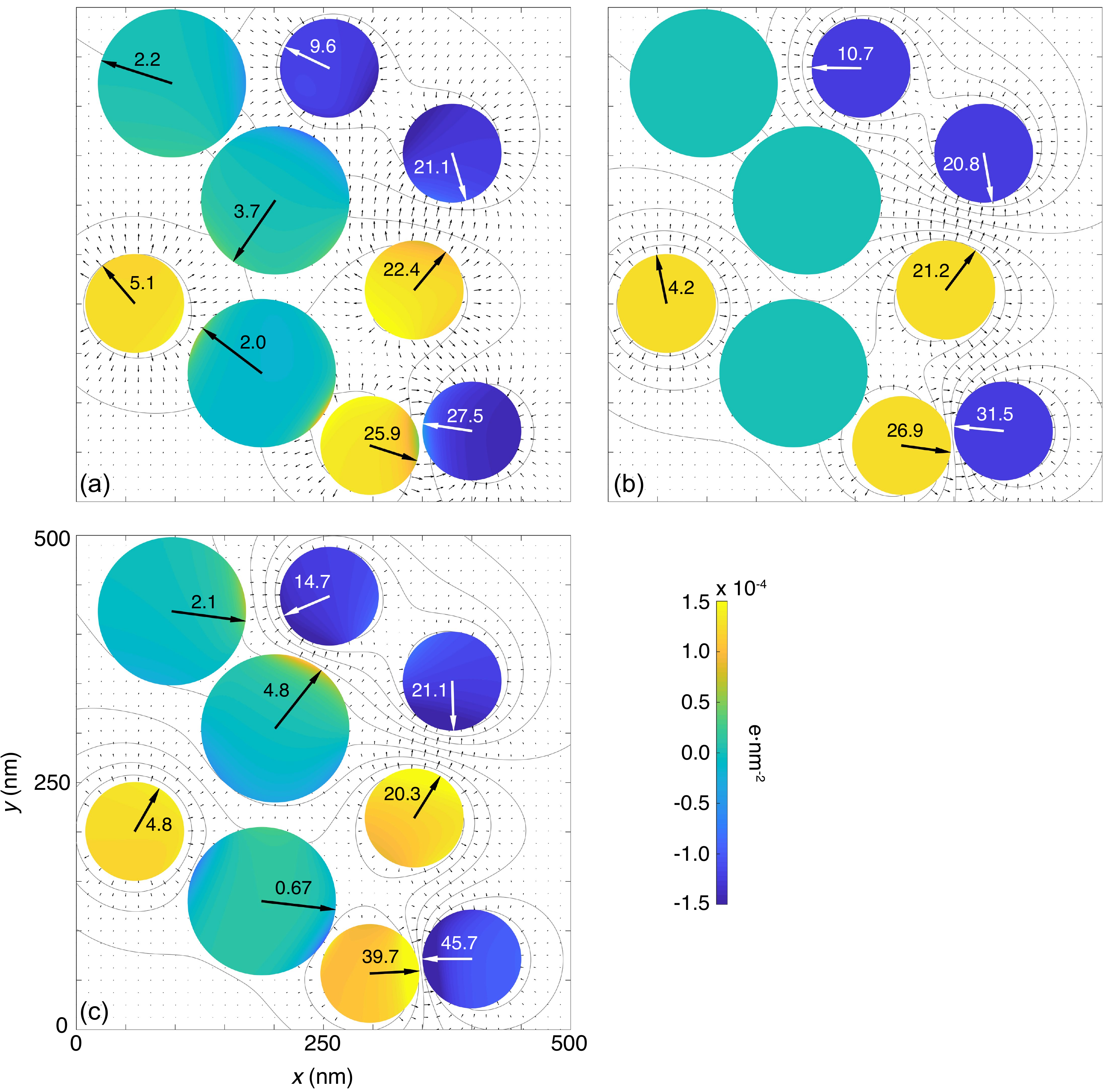}
	\caption{Schematic illustration of the electrostatic interaction in a system made by three representatives of positive, negative and neutral particles, randomly displaced in pure water ($\varepsilon_\text{m} = 80$; [1:1 electrolyte] = 10$^{-7}$ molar), for three cases of particle dielectric constant: $\varepsilon_i = 20$ (a), $\varepsilon_i = 80$ (b), and $\varepsilon_i = 320$ (c), where $i=1,2,\dots,9$. Depicted are the forces (with directions denoted by the arrows and magnitudes denoted by the accompanying numbers) on each particle, equipotential lines and electric field vectors. The colors of the illustrative particles reflect the sign and magnitude of the total surface charge density, from deep blue (most negative) to deep yellow (most positive). Naturally, such variation of density is more evident on the neutral particle.}
	\label{fig:Force107}
\end{figure}

In the first scenario, the medium is pure water, i.e. with $\varepsilon_\text{m} = 80$ and an ionic strength caused by an 1:1 electrolyte concentration of 10$^{-7}$ M, and all particles have a particular value of dielectric constant in three separate occasions.
As depicted in Figure \ref{fig:Force107}, with all other parameters of the system being fixed, a change of particle dielectric constant, from $\varepsilon_i = 20$, passing through $\varepsilon_i = 80$ and on to $\varepsilon_i = 320$ ($i=1,2,\dots,9$), can have a significant influence on the magnitude of each particule force and especially in its direction, with such influence being more noticeable for the neutral particles.
Accordingly, it can be noted that for the scenario depicted in this figure, the interactions between charged particles are dominated by the Coulomb (monopole) component of the electrostatic force, with some contribution advent from polarization effects when $\varepsilon_i \neq \varepsilon_\text{m}$, which effectively provoke a modulation of the net electrostatic force. For instance, the force between opposite charged particles becomes more attractive when $\varepsilon_i > \varepsilon_\text{m}$ and less attractive when $\varepsilon_i < \varepsilon_\text{m}$; on the other hand, the force between like-charged particles becomes less repulsive when $\varepsilon_i > \varepsilon_\text{m}$ while more repulsive when $\varepsilon_i < \varepsilon_\text{m}$. However, strictly speaking, in a many-body system one cannot refer to an interaction between any two particles alone; it just turns out that the net force acting on charged particles in this particular system is dominated by the Coulomb component, fact that is not true for the three neutral particles present in the system. In fact, the forces acting on the latter are entirely originated from polarization. 
Accordingly, when $\varepsilon_i \neq \varepsilon_\text{m}$, the neutral particles are subject to the electric field created by the charged species of the system, leading to anisotropic creation of bound charges at their surfaces. These bound charges also participate in the process of mutual polarization, leading to an overall equilibrium distribution of charges involving, in a coupled manner, all particles of the system. The practical effect is a dramatic change in the forces acting on the neutral particles, which flips direction from $\varepsilon_i < \varepsilon_\text{m}$ to $\varepsilon_i > \varepsilon_\text{m}$, while changes in the direction of the forces acting on the charged particles are more subtle since, as aforementioned, these are dominated by Coulomb component. 
At $\varepsilon_i = \varepsilon_\text{m}$, the lack of particle polarization ensures no electrostatic force acts on the neutral particles, and in such case the forces acting on the charged particles could be then described by the electrostatic component of DLVO theory. 

\begin{figure}[h]
	\centering
	\includegraphics[width=0.72\textwidth]{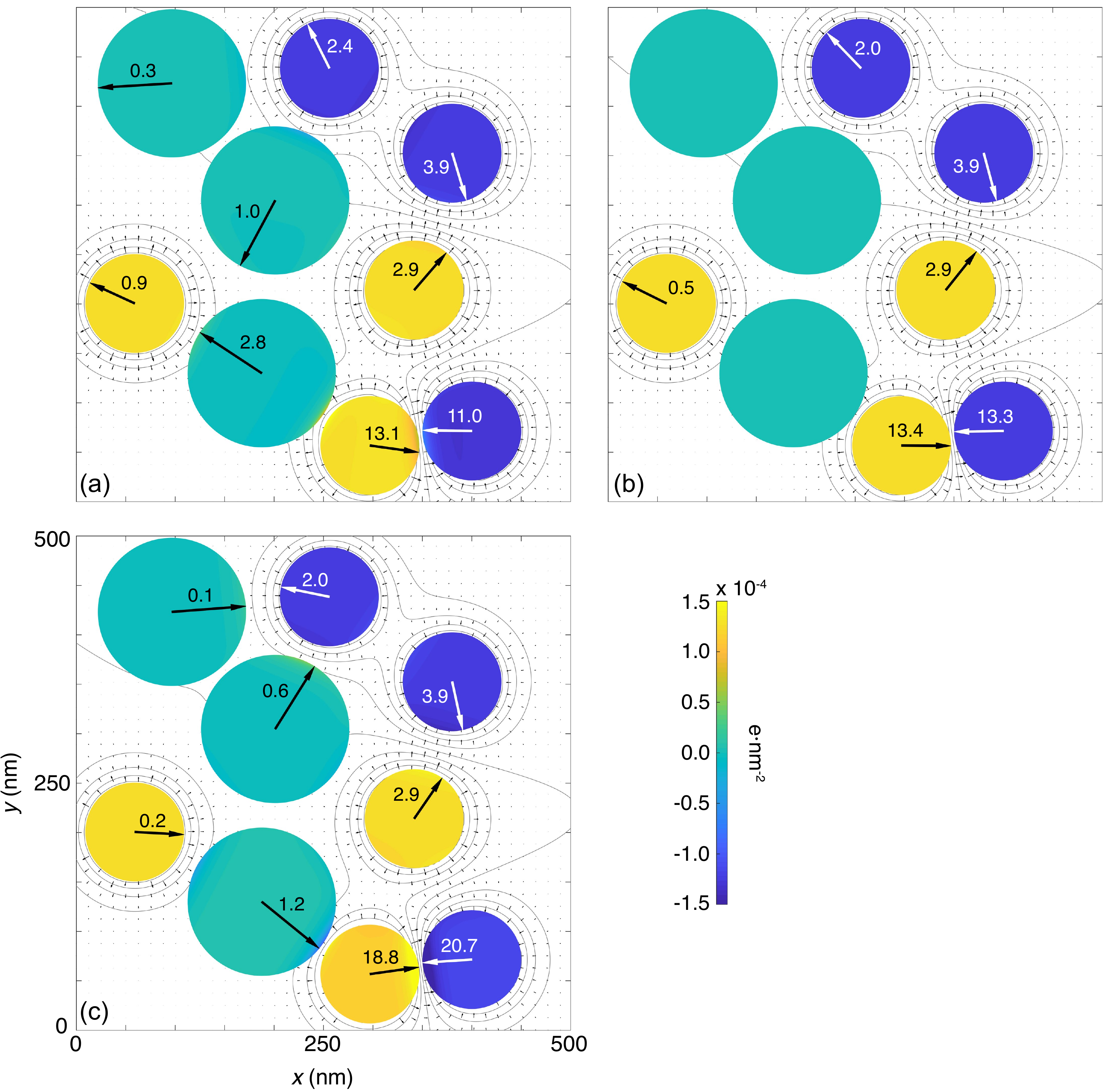}
	\caption{Schematic illustration of the electrostatic interaction in the same system addressed at Figure \ref{fig:Force107} but with a higher electrolyte concentration (10$^{-4}$ molar), for three cases of particle dielectric constant: $\varepsilon_i = 20$ (a), $\varepsilon_i = 80$ (b), and $\varepsilon_i = 320$ (c), where $i=1,2,\dots,9$.}
	\label{fig:Force104}
\end{figure}

In a second scenario, the system remains the same, apart from the aqueous medium that is added by an 1:1 electrolyte, rendering an ionic concentration of 10$^{-4}$ molar. As can be noted from Figure \ref{fig:Force104}, the higher ionic strength screens more effectively the electrostatic interactions, generally rendering the forces and polarization effects weaker. Such results are also pictured graphically by equipotential lines and electric field vectors (whose sizes represent the field magnitude), also shown in each of the figures. As can be seen, the neighboring equipotential lines are less distanced from one another in Figure \ref{fig:Force104} ($\kappa^{-1} = 30$ nm) than in Figure \ref{fig:Force107} ($\kappa^{-1} \approx$ 1 $\mu$m), reflecting the faster decay of the electric potential in the former case. 
Further, it can also be observed that the arrows representing the electric field at points across the system are visibly more restricted to regions around the charged particles in Figure \ref{fig:Force104}, while in Figure \ref{fig:Force107} the arrows are seen more widely distributed through the plotted area, also reflecting the more gradual decay of the potential in the latter, since $\textbf{E} = - \grad \Phi$.

The net electrostatic force on the neutral particle that is closest to the bottom provides a clear illustration of how coupled many-body and screening effects are capable of modulating an interaction. In Figure \ref{fig:Force107}, screening effects are minimum ($\kappa^{-1} \approx$ 1 $\mu$m) and the force on such neutral particle is a consequence of polarization mostly exerted by the three positive particles; one situated at southeast with a surface-to-surface separation $s = 6.6$ nm, other situated at northwest with $s = 21.6$ nm, and another at northeast with $s = 50.6$ nm. Taking as an example the case where all particles are more polarizable than the medium, namely Figure \ref{fig:Force107}(c), it can be noted that the magnitude of the force acting on the neutral particle (0.67 pN) is particularly small, due to a balance of influences. Accordingly, since the two closest positive particles are in opposite positions in relation to the neutral one, the net force acting on the latter is slightly in favor to the closer particle at southeast. In addition, as screening effects are minimum, the more distanced particle at northeast can also exert its influence, slightly deflecting in its favor the direction of the force acting on the neutral particle.

The situation changes with the addition of the electrolyte, as illustrated in Figure \ref{fig:Force104}, where screening effects are more significant ($\kappa^{-1} \approx$ 30 nm), effectively changing the previous balance of influences. Accordingly, the positive particle at northeast is now at a surface-to-surface distance from the neutral particle that is greater than the Debye length of the solution, and the same quantity between the particle at northwest and the neutral particle is of approximately $\nicefrac{2}{3} \; \kappa^{-1}$, allowing this positive particle to still exert a minor influence over the neutral. Therefore, being at a surface-to-surface separation of approximately just $\nicefrac{1}{3} \; \kappa^{-1}$, the particle at southeast can exert more dominantly its influence over the neutral species, which experience a net force of greater magnitude (1.2 pN) and that is completely directed to the former.

\subsection*{Electrostatic and van der Waals interactions}

While the proposed model provides an accurate description of electrostatic interactions between dielectric particles by also accounting for mutual particle polarization---effect that is particularly important at short inter-particle separation distances---the always present van der Waals (vdW) forces must also be considered in a more complete description of a dielectric system, as such forces are largely insensitive to electrolyte concentrations, and notably dominant at even shorter separation distances (closer to the particle's touching point). \cite{Israelachvili} 

In this regard, the balance between electrostatic (usually without polarization) and vdW forces is well known for being a critical factor to the kinetic nonlability of colloids. Accordingly, colloidal particles usually bear like-charges at their surfaces and therefore are able to stay well stably dispersed in solution on the account of electrostatic repulsion. With an increasing ionic strength of a colloidal solution, the clustering of ions of opposite charge at regions around the particles becomes more significant, causing a decrease to the value of $\kappa^{-1}$ (formally defined as the characteristic thickness of the so-called electrical double layer), which consequently translates to a more effective screening of the repulsive electrostatic forces, therefore possibly extending to longer lengths the influence exerted by the short-range vdW forces. 

In sufficiently strong electrolyte solutions, the screening is so effective that a secondary (and local) minimum can appear in the energy surface of interacting colloidal particles, as will be featured in the upcoming calculations. Coagulation occurs when the particles are able to reach the primary (and global) energy minimum, situated at the touching point, where they are thermodynamically stable. However, if the particles are sufficiently charged, the electrostatic repulsive barrier might be high enough to prevent coagulation, with the particles ending up dispersed in solution, if no secondary minimum is present or, otherwise, undergoing flocculation by staying weakly aggregated in a energy well situated before the repulsive barrier, generally at a few $\kappa^{-1}$ from the touching point, rendered by a balance between electrostatic and vdW forces.

Such phenomenon is commonly accounted by the DLVO theory, that combine contributions from the electrical double layer and vdW forces, but in such case the possible contributions made by polarization effects are not addressed. Calculations were then performed to explore how an additional component to the total interaction, introduced by the account of particle polarization, might affect the stability of colloidal particles. The system of interest is composed by a given number of identical particles, each with 50 nm-radius and a surface charge $\sigma_f = 0.3$ e$\cdot$nm$^{-2}$, immersed in a aqueous medium $(\varepsilon_\text{m}= 80)$ with $\kappa^{-1} \approx$ 1 nm, rendered by $10^{-1}$ molar of a 1:1 electrolyte. The geometry consists of one of the particles fixed at a centered position and surrounded by a layer formed by 4, 8 or 12 particles, all at the same distance from the centered particle. Those that form the layer are displaced in a way that minimizes the electrostatic energy within the layer when in contact with the centered particle. The total potential energy, relative to the sum of electrostatic and vdW interactions, is then traced as function of the surface-to-surface separation, $ s $, between the central particle and the particles that form the layer, which is effectively accomplished by radially and simultaneously moving each of the latter outwards. The dielectric constants are piecewise constant and inside $\Oi$ three possible values can be chosen, $\varepsilon_i = 20, 80, 320$, $i=1,2,\ldots,M$, allowing then the effects of having particles that are less, equal or more polarizable than the medium to be addressed. A Hamaker constant equal to $ 10^{-19} $ J was considered for the calculation of van der Walls interactions, and these were approximated by using the pairwise expression given by Israelachvili \cite{Israelachvili} that is valid at all inter-particle separations.

\begin{figure}[h]
	\centering
	\includegraphics[width=0.6\textwidth]{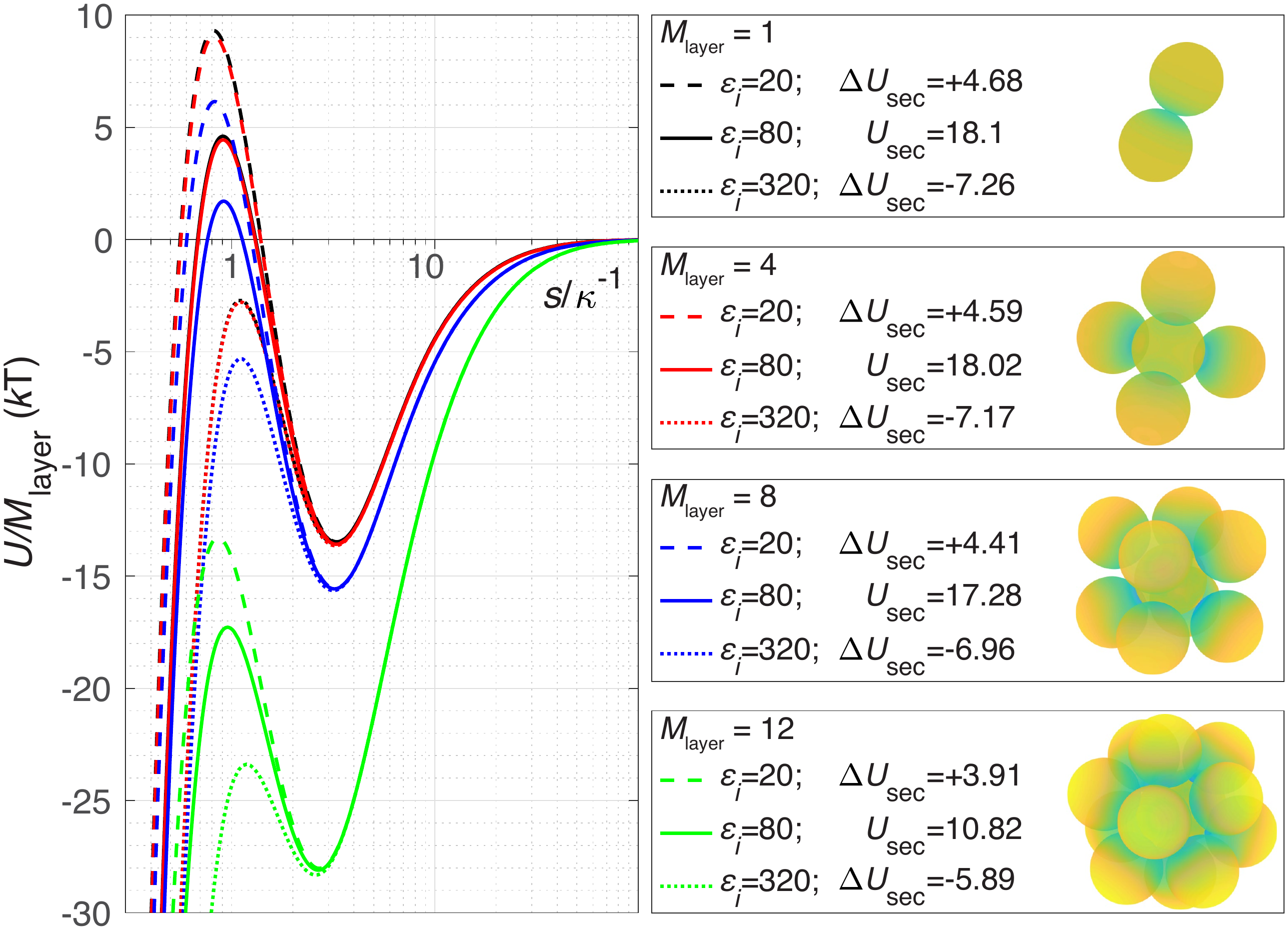}
	\caption{Plot of the electrostatic potential energy weighted by the number of layer forming particles, $ U/M_{\text{layer}} $, as function of the shared surface-to-surface separation between layer forming particle and the centered particle, weighted by the Debye length, $ s/\kappa^{-1} $. $ U_{\text{sec}} $ denotes the height of the energy barrier, i.e. the difference between its maximum and the secondary minimum, for the case where $ \varepsilon_i = 80 $, and $ \Delta U_{\text{sec}} $ denotes the corresponding relative difference in energy that occurs when $ \varepsilon_i = 20 $ and $ \varepsilon_i = 320 $.}
	\label{fig:Energy}
\end{figure}

Observations that are shared between the different geometries can be extracted from the results, presented in Figure \ref{fig:Energy}. Accordingly, the secondary minimum is always present and is located at a separation of a few Debye lengths from the touching point. For each case, the curves representing different particle dielectric constants starts to diverge approximately when the distance is of three Debye lengths. Towards shorter separations, the energy barriers to coagulation ascend pronouncedly, reaching a maximum with a magnitude and location that is characteristic for each value of the particles' dielectric constants; correspondingly, the transition from a particle dielectric constant of 320 to 20 causes an increase in the magnitude of the barrier and shifts its location towards the touching point.

Taking the case where $ M_{\text{layer}} = 12 $ as an example, in the absence of particle polarization, i.e. when $\varepsilon_i = \varepsilon_\text{m} = 80$ ($i=1,2,\ldots,12$), the energy barrier, weighted by the number of particles that form the layer, has a height of 10.82 $kT$ and its peak is located at a distance of $0.96 \, \kappa^{-1}$ from the touching point ($ s=0 $). A value of particle dielectric constant that is four times smaller than that of the medium provokes a significant increase of 3.91 $kT$ to the energy barrier, which becomes about 36.1\% higher, and shifts the location of its peak to $0.86 \, \kappa^{-1}$; as the particles are less polarizable than the medium, an effective reverse polarization takes place, consequently provoking a strengthening of repulsive electrostatic forces between the particles. Going to the other extreme, a value of particle dielectric constant that is four times greater than that of the medium provokes a decrease of the energy barrier to 4.93 $kT$, almost half of its original value, and shifts the location of its peak to $1.18 \, \kappa^{-1}$; as the particles are now more polarizable than the medium, polarization effects cause a weakening of the electrostatic force between the particles, which becomes less repulsive.

Twelve is the maximum number of spherical particles that can be arranged around and at a same distance from a central same-sized particle without incurring in superposition of volumes at the minimum considered distance, $ s = 0 $. Even though a significant electrostatic shielding is produced by the electrolyte solution, when $ s \rightarrow 0 $ and $ M_{\text{layer}} = 12 $, the particles in the layer are able to strongly interact not only with particle in the center but also with their nearest neighbors within the layer. As illustrated in Figure \ref{fig:Energy}, such particles assume positions that can be closely translated to the vertices of a regular icosahedron. The geometry changes to a cube and a tetrahedron when the number of particles that form the layer decreases to eight and four, respectively. It happens that the particles in the layer are more spaced in the last two geometries which, coupled with the screening provided by the solution, causes them to interact less strongly with one another. This is reflected on their potential energy curves. As illustrated in Figure \ref{fig:Energy}, there are very small differences between curves for the cases where $ M_{\text{layer}} = 4 $ and $ M_{\text{layer}} = 1 $ (the latter being effectively a two-body interaction). Accordingly, the layer forming particles in the former case are so distanced from one another that the electrostatic screening effects turn their interactions to virtually zero and, consequently, the energy weighted by the number of particles in the layer approaches that of a two-body interaction ($ M_{\text{layer}} = 1 $), indicating then that the overall interaction when $ M_{\text{layer}} = 4 $ is mostly dictated by the interaction between each particle in the layer with the centered particle. As can be also noted, the presence of more particles in the layer causes a displacement of the energy curves, normalized with respect to the number of particles that form the layer, towards lower values. 
This effect is already noticeable when $ M_{\text{layer}} = 8 $ and especially evident when $ M_{\text{layer}} = 12 $, where the layer forming particles are sufficiently close to each other provided that $ s \rightarrow 0 $, which allows them to interact strongly, despite the substantial screening effects. 
Geometric considerations apart, particle polarization also influences the energy barriers in all cases, in a similar way to that earlier described for $ M_{\text{layer}} = 12 $. For instance, such effects causes the weighted energy barrier to become 25.5\% higher when $ M_{\text{layer}} = 8 $ and $\varepsilon_i = 20 $ ($ i=1,2,\ldots,9 $), or 39.8\% smaller when $ M_{\text{layer}} = 4 $ and $\varepsilon_i = 320 $ ($ i=1,2,\ldots,5 $).

It is evident from the presented results that polarization effects can have a significant influence on the interaction between charged dielectric particles, even in the presence of weak or strong screening agents, if the distance between interacting particles is sufficiently small. The results suggest that such a ``sufficiently small'' distance can be translated to one that is comparable or smaller than the Debye length, as to allow a particle to experience a sufficiently strong external electric field (as, in the presented cases, generated by charges on other particles) and become significantly polarized.

\section*{Conclusions}

A general many-body model for the problem of calculating electrostatic interactions between charged dielectric particles, in the presence of a medium that can be additionally permeated by charged carriers, has been presented. The model is able to accurately describe the mutual polarization experienced by interacting dielectric particles, which constitutes a true many-body problem if three or more particles are involved, as such kind of interaction can be highly nonadditive. The inherent ability of the model to determine a maximum degree of spherical harmonics, in which its discretization is based upon, allows a fine and systematic control of the accuracy of results, by permitting the consideration of high order induced multipole in each particle, well beyond the usual dipole, quadrupole or, more rarely, octopoles, commonly employed by other methods in the literature. 

The additional presence of charge carries in the medium can act to screen particle interactions and weaken polarization effects. Notwithstanding, it has been shown that even in strong electrolyte solutions, particles can still undergo significant polarization and therefore deviate from a pure Coulombic interaction, if their separations is comparable to or smaller than the Debye length of the medium.  Such deviation from pure Coulombic interactions can be substantial in some situations, and even result in a complete reversal of the acting electrostatic force on neutral but polarizable particles, or extensive modulation of the energy barrier related to colloidal solutions.

On its potential of applicability in describing experimental results, the model has significance not only to cases where particles of dielectric nature are moderately or strongly polarizable, as would be the case for interactions in vacuum but, in fact, also to scenarios where there exist a dielectric contrast between the particles (herein denoted as $ \Omega_i $, where $ i=1,2,\ldots,M $) and the medium (herein denoted as $ \Omega_0 $). For instance, in the context of colloidal solutions, it is recognized that the permeating medium is commonly water, which has a dielectric constant of approximately 80 at room temperature. Then, unless the colloidal particles are polarizable to the same extent as water, additional multipolar forces will also operate in the system. Indeed, colloidal particles frequently have a dielectric constant that is smaller than that of water so, according to the obtained results, the true energy barrier in such cases will be higher than that predicted by descriptions that do not take into account polarization effects, such as the commonly employed DLVO theory, and therefore underestimated.

\section*{acknowledgements}
Chaoyu Quan acknowledges the funding from the PICS-CNRS and the PHC PROCOPE 2017 (Project No. 37855ZK).
Benjamin Stamm acknowledges the funding from the German Academic Exchange Service (DAAD) from funds of the “Bundesministeriums für Bildung und Forschung” (BMBF) for the project Aa-Par-T (\text{Project-ID 57317909}).

\section*{Appendix A}
\setcounter{equation}{0}
\renewcommand\theequation{A\arabic{equation}}

As a generalization and combination of the works of Lindgren et al. \cite{LindgrenManyBody2018} and Quan et al. \cite{quan2018domain}, an integral equation formulation can be derived for the problem described by the partial differential equations~(\ref{eq:PDE_out})-(\ref{eq:BC2}) with the initially unknown potential $\Phi$.
Indeed, following the arguments presented in Lindgren et al. \cite{LindgrenManyBody2018} but with an ionic concentration, i.e. for $\kappa>0$, the restriction $\lambda=\Phi|_{\Gamma_0}$ is defined by the following integral equation on $\Gamma_0 = \Gamma_1\cup \ldots \cup \Gamma_M$:

\begin{equation}
	\label{eq:IntegralEquation}
	\left(I_\text{id} - \mathcal{S}_\kappa \mathcal{L}_\kappa \right) \lambda = \dfrac{K}{\varepsilon_\text{m}} \mathcal{S}_\kappa \sigma_f,
\end{equation}
where $\mathcal S_\kappa$ denotes the single layer boundary operator
\[
\forall x \in \Gi: \quad \left( \mathcal{S}_\kappa g \right) (x) 
=
\int_{\Gamma_0} \frac{g(y)}{|x-y|} \, e^{-\kappa |x-y|} \, dy,
\]
and on each $\Gi$, $\mathcal{L}_\kappa|_{\Gi} = \mathcal{L}_\kappa^i$ is a local operator given by
\[
\forall x \in \Gi: \quad (\mathcal{L}_\kappa^i \, g^i) (x) = \left( \text{DtN}_\kappa^i (g^i) - \dfrac{\varepsilon_i}{\varepsilon_\text{m}} \text{DtN}_0^i (g^i) \right) (x) .
\]
Here, $\text{DtN}_\kappa^i$ denotes the Dirichlet-to-Neumann map associated to the operator $-\Delta+\kappa^2$ in the ball $\Omega_i$. Further details will be given below.

A numerical method is proposed to discretized \eqref{eq:IntegralEquation} on $\Gamma_0$, where the Galerkin method is used with a series of real spherical harmonics on each sphere as basis functions. 
To be precise, a numerical approximation  $\lambda_N$ of $\lambda$ on each $\Gj$ is introduced as follows
\begin{equation}
	\label{eq:lambda}
	\lambda_{N,j} (x) = \lambda_N |_{\Gj} (x) = \sum_{\ell^\prime=0}^{N} \sum_{m^\prime =-\ell^\prime}^{\ell^\prime} \left[ \lambda_j    \right]_{\ell^\prime}^{m^\prime} \mathcal Y_{\ell^\prime m^\prime} \!\left( \frac{x-x_j}{|x-x_j|}\right), \quad \forall x \in \Gj.
\end{equation}
Here, $\Ylm$  represents the real spherical harmonic function of degree $\ell$ and order $m$.

For each $\Gj$, (\ref{eq:IntegralEquation}) is multiplied by a test function $\Ylm \left( \frac{x-x_i}{|x-x_i|}\right)$ on the both sides and then integrated over $\Gi$, to obtain the weak formula: find $\lambda_N$ in the form of (\ref{eq:lambda}) such that for all $j=1,\ldots,M$ and $\ell,m$ with $\ell=0,\ldots,N$, $-\ell \leq m\leq \ell$ there holds
\[
\int_{\Gi} \Ylm \! \left( \frac{x-x_i}{|x-x_i|}\right) \left(I_\text{id} - \mathcal{S}_\kappa  \!  \mathcal{L}_\kappa \right) \lambda_N(x)  \, dx
= 
\dfrac{1}{\varepsilon_\text{m}} \int_{\Gi} \!\left( \mathcal{S}_\kappa \sigma_f \right)(x) \, \Ylm \!\left( \frac{x-x_i}{|x-x_i|}\right) \, dx.
\]
In addition, according to \eqref{eq:sigma_f}, $\sigma_f$ can be written as
\[
\forall x \in \Gi: \quad \sigma_f |_{\Gi} (x) = \left[ \sigma_{f,i} \right]_0^0 \Yzz ,
\]
where $\Yzz$ is constant.

The involved integrals cannot always be integrated analytically. 
As a consequence, for each $\Gamma_i$, the following Lebedev quadrature rule is used:
\[
\left\langle f, g \right\rangle _{i, N_{\rm leb}} = \sum_{n=1}^{N_{\rm leb}} \omega_n f(x_i + r_i s_n) g(x_i + r_i s_n),
\]
\noindent where $N_{\rm leb}$ is the number of Lebedev points, $w_n \in \mathbb{R}$ and $s_n \in \mathbb{S}^2$ are Lebedev integration weights and points on the unit sphere. 
This quadrature provides an approximation to the integral of $fg$ over $\Gamma_i$, that is, 
\[
\left\langle f, g \right\rangle _{i,N_{\rm leb}} \approx  \int_{\mathbb{S}^2} f(x_i + r_i \hat{s}) g(x_i + r_i \hat{s}) d\hat{s} = \dfrac{1}{r_i^2} \int_{\Gi} f(s) g(s) ds.
\]
\noindent Note that if $f$ and $g$ are spherical harmonics up to a certain degree, this approximation is exact when a sufficiently large number of integration points is chosen.
With the Lebedev quadrature rule, the problem is then written as: find $\lambda_N$ in the form of (\ref{eq:lambda}) such that
\[
\forall_i, \forall_{\ell,m}: \quad \left\langle \Ylm \left( \frac{\cdot-x_i}{|\cdot-x_i|}\right), \left(I_\text{id} - \mathcal{S}_\kappa \mathcal{L}_\kappa \right) \lambda_N \right\rangle _{i,N_{\rm leb}} = \dfrac{1}{\varepsilon_\text{m}} \left\langle \Ylm \left( \frac{\cdot-x_i}{|\cdot-x_i|}\right), \mathcal{S}_\kappa \sigma_f \right\rangle _{i,N_{\rm leb}},
\]
where, again, only $\ell=0,\ldots,N$ is considered.
By linearity, this can be expressed in terms of a linear system
\[
A {\lambdav} ={\bm f}
\]
\noindent with
\begin{align*}
	\left[ A_{ij} \right] _{\ell\ell^\prime} ^{mm^\prime} = & \left\langle \Ylm \left( \frac{\cdot-x_i}{|\cdot-x_i|}\right), \left(I_\text{id} - \mathcal{S}_\kappa \mathcal{L}_\kappa^j \right) \Ylpmp \left( \frac{\cdot-x_j}{|\cdot-x_j|}\right) \right\rangle _{i,N_{\rm leb}} , \\
	\left[ f_{i} \right] _{\ell} ^{m} = & \dfrac{1}{\varepsilon_\text{m}} \left\langle \Ylm \left( \frac{\cdot-x_i}{|\cdot-x_i|}\right), \mathcal{S}_\kappa \sigma_f \right\rangle _{i,N_{\rm leb}}.
\end{align*}
In order to provide explicit expressions of the matrix elements, we should study carefully the two operators $\mathcal{L}_\kappa^j$ and $\mathcal{S}_\kappa$ in the following content.

We first consider the local operator $\mathcal{L}_\kappa^j$.
In the sphere $\Omega_j=B_r(x_j)$, the related local problem is 
\begin{equation*}
	\label{eq:SingleSphPb}
	\begin{cases} 
		(-\Delta + \kappa^2) u_j = 0 & \text{in} \; \Oj = B_r(x_j), \\
		\qquad\qquad\;\; \, u_j = g_j & \text{on} \; \Gj = \partial \Oj, 
	\end{cases}
\end{equation*}
for a given local Dirichlet trace $g_j$ which is supposed to be given as an expansion of spherical harmonics as follows
\begin{equation}
	\label{eq:gdef}
	g_j(x) = \sum_{\ell=0}^{N} \sum_{m=-\ell}^\ell [g_j]_\ell^m \, \Ylm\left( \frac{x-x_j}{|x-x_j|}\right), \quad \forall x \in \Gj.
\end{equation}
\noindent Then, $u_j$ is given by 
\[
u_j(x) = \sum_{\ell=0}^{N} \sum_{m=-\ell}^\ell [g_j]_\ell^m \,\dfrac{f_\ell^\kappa (|x-x_j|)}{f_\ell^\kappa (r_j)} \, \Ylm\left( \frac{x-x_j}{|x-x_j|}\right), \quad \forall x \in \Oj,
\]
with 
\[
f_\ell^\kappa (r) = i_\ell (\kappa r) \quad \text{and} \quad (f_\ell^{\kappa})^\prime (r) = \kappa \, i_\ell^\prime(\kappa r) \text{,} 
\]
\noindent where $ i_\ell $ is the modified spherical Bessel function of the first kind (see Quan et al. \cite{quan2018domain}). Note that the limit $\kappa \to 0$ yields the case of the Laplace equation and $f_\ell^0$ becomes simply the polynomial
\[
f_\ell^0 (r) = r^\ell \quad \text{so that} \quad (f_\ell^{0})^\prime (r) = \ell \, r^{\ell-1} \text{.} 
\]
\noindent The normal derivative of $u_j$ at $\Gamma_j$, thus the Dirichlet-to-Neumann map, then writes for all $x\in\Gj$:
\begin{align*}
	\text{DtN}_\kappa^j (g_j) (x)
	= \grad u_j \cdot n (x) 
	&= \sum_{\ell=0}^{N} \sum_{m=-\ell}^\ell [g_j]_\ell^m \,\dfrac{(f_\ell^{\kappa})^\prime (|x-x_j|)}{f_\ell^\kappa (r_j)} \, \Ylm \left. \left( \frac{x-x_j}{|x_0-x_j|}\right) \right|_{x \in \Gj},\\
	&= \sum_{\ell=0}^{N} \sum_{m=-\ell}^\ell [g_j]_\ell^m \,\dfrac{(f_\ell^{\kappa})^\prime (r_j)}{f_\ell^\kappa (r_j)} \, \Ylm \left( \frac{x-x_j}{r_j}\right),\\
	&= \sum_{\ell=0}^{N} \sum_{m=-\ell}^\ell [g_j]_\ell^m \, [z_j^\kappa]_\ell \, \Ylm  \left( \frac{x-x_j}{r_j}\right) ,
\end{align*}
\noindent with
\[
[z_j^\kappa]_\ell = \dfrac{(f_\ell^{\kappa})^\prime (r_j)}{f_\ell^\kappa (r_j)}
\quad\mbox{and}\quad
[z_j^0]_\ell = \dfrac{\ell}{r_j}.
\]
\noindent Then, for $g_j$ of the form given in \eqref{eq:gdef}, one obtains on $\Gamma_j$:
\begin{align*}
	\forall x \in \Gj: \quad (\mathcal{L}^j_\kappa \, g_j) (x) &= \left( \text{DtN}_\kappa^j (g_j) - \dfrac{\varepsilon_j}{\varepsilon_\text{m}} \text{DtN}_0^j (g_j) \right) (x),\\
	&= \sum_{\ell=0}^{N} \sum_{m=-\ell}^\ell [g_j]_\ell^m \, \left( [z_j^\kappa]_\ell - \dfrac{\varepsilon_j}{\varepsilon_\text{m}} [z_j^0]_\ell \right) \, \Ylm\left( \frac{x-x_j}{r_j}\right),\\
	&= \sum_{\ell=0}^{N} \sum_{m=-\ell}^\ell [g_j]_\ell^m \,[L_j]_\ell \, \Ylm\left( \frac{x-x_j}{r_j}\right),
\end{align*}
\noindent with
\[
[L_j]_\ell = [z_j^\kappa]_\ell - \dfrac{\varepsilon_j}{\varepsilon_\text{m}} [z_j^0]_\ell,
\]
and thus
\[
\forall x \in \Gj: \quad \left(\mathcal{L}^j_\kappa \, \Ylm \! \left( \frac{\cdot-x_j}{|\cdot-x_j|}\right) \right) (x)
=
[L_j]_{\ell} \, \Ylm \!\left( \frac{x-x_j}{|x-x_j|}\right).
\]

Next, we consider the single layer boundary operator $\mathcal S_\kappa$ appearing in the matrix elements.
Let $h_j (x) = \sum_{\ell=0}^{N} \sum_{m=-\ell}^\ell [h_j]_\ell^m \, \Ylm\left( \frac{x-x_j}{|x-x_j|}\right)$, with $[h_j]_\ell^m = [g_j]_\ell^m [L_j]_\ell$, be given on each $\Gamma_i$ as developed above, which corresponds to $h_j=\mathcal{L}_\kappa^j \, g_j$ on $\Gamma_j$. 
Then, the potential $\mathcal{S}_\kappa h_j$ on $\Gamma_i$ can be computed as follows
\[
\forall x \in \Gi: \quad \left( \mathcal{S}_\kappa h_j \right) (x) = \sum_{j=1}^M \sum_{\ell=0}^{N} \sum_{m=-\ell}^\ell [h_j]_\ell^m [n_j]_\ell \, \dfrac{k_\ell \left(\kappa|x-x_j|\right)}{k_\ell(\kappa \, r_j)} \, \Ylm\!\left( \frac{x-x_j}{|x-x_j|}\right),
\]
\noindent with
\[
[n_j]_\ell = \left( \kappa \dfrac{i_\ell^\prime (\kappa \, r_j)}{i_\ell (\kappa \, r_j)} - \kappa \dfrac{k_\ell^\prime (\kappa \, r_j)}{k_\ell (\kappa \, r_j)} \right)^{-1},
\]
\noindent where $ k_\ell $ is the modified spherical Bessel function of the second kind.
One can refer to Quan et al. \cite{quan2018domain} for details.

For the elements of the matrix $A$, note first that if $i=j$, the exact integration can be obtained as follows
\begin{equation}
	\left[ A_{ii} \right] _{\ell\ell^\prime} ^{mm^\prime} = \delta_{\ell\ell^\prime} \delta_{mm^\prime} \left( 1 - \left[n_i\right]_\ell \left[L_i\right]_\ell \right),\label{eq:Aii}
\end{equation}
\noindent otherwise, we have
\begin{equation}
	\left[ A_{ij} \right] _{\ell\ell^\prime} ^{mm^\prime} = - \sum_{n} \Ylm(s_n) \, \omega_n \left[n_j\right]_{\ell^\prime} \left[L_j\right]_{\ell^\prime} \dfrac{k_{\ell^\prime} \left(\kappa r_n^{ij}\right) }{ k_{\ell^\prime} \left(\kappa r_j\right) } \mathcal Y_{\ell^\prime m^\prime} (s_n^{ij}),\label{eq:Aij}
\end{equation}
\noindent where
\[
s_n^{ij} = \dfrac{x_i + r_i s_n - x_j}{\left|x_i + r_i s_n - x_j\right|} \quad \mbox{and} \quad r_n^{ij} = \left|x_i + r_i s_n - x_j\right|.
\]
\noindent In addition, it holds for the column vector $\mathbf f$ that
\[
\left[f_i\right] _\ell ^m = \dfrac{K}{\varepsilon_\text{m}} \sum_{n=1}^{N_{\rm leb}} \Ylm(s_n) \, \omega_n \sum_{j=1}^M \left[n_j\right]_0 \dfrac{k_0 \left(\kappa r_n^{ij}\right) }{ k_0 \left(\kappa r_j\right) } \Yzz \left[\sigma_{f,j}\right] _0 ^0.
\]

\subsection*{1. Electrostatic energy}\label{sect:elec_eng}
As demonstrated in Lindgren et al. \cite{LindgrenManyBody2018}, the total electrostatic energy of the system, $ U $, has a discrete approximation as follows
\begin{align}
	\label{eq:EnergyIntRepDisc}
	U(\lambda_N,\sigma_f)
	=
	\frac{1}{2}
	\sum_{i=1}^M
	\int_{\Gi} \sigma_{f,i}(s) \lambda_{N,i}(s) ds
	=
	\langle \Psiv , \lambdav \rangle,
\end{align}
where the entries of vector $\Psiv$ are given by
\[
[\Psi_i]_\ell^m 
= 
\delta_{\ell 0} \frac{r_i^2}{2} [\sigma_{f,i}]_0^0,
\]
and
\[
\langle \Psiv , \lambdav \rangle 
\coloneqq
\sum_{i=1}^M  \sum_{\ell=0}^N \sum_{m=-\ell}^\ell  [\Psi_i]_\ell^m [\lambda_i]_\ell^m
=
\sum_{i=1}^M \frac{r_i^2}{2} [\sigma_{f,i}]_0^0 [\lambda_i]_0^0.
\]

The self-energies can be obtained by solving \eqref{eq:PDE_out}--\eqref{eq:BC2} for each individual particle $\Oi$, while neglecting the presence of the other spheres in the system. 
Accordingly, the solution on $\Gi$, denoted here by $\lambda_{f,i}$, is given by
\[
\lambda_{f,i} = \frac{K}{k_0} \SLO_i \sigma_{f,i}.
\]
Thus, the total self-energy $U^{\rm self}$ of the system, i.e., the sum of the self-energy of each individual particle in isolation, is obtained as follows
\begin{align}
	\label{eq:EnergyIntRepDiscSelf}
	U^{\rm self}(\sigma_{f})
	=
	\frac{1}{2}
	\sum_{i=1}^M
	\int_{\Gi} \sigma_{f,i}(s) \lambda_{f,i}(s) ds
	=
	\langle \Psiv , \lambdav_{f} \rangle, 
\end{align}
where the entries for the vector $\lambdav_{f}$ are given by
\[
[\lambda_{f,i}]_\ell^m 
= 
\delta_{\ell 0}  \frac{4\pi r_i K}{k_0} [\sigma_{f,i}]_0^0.
\]

The energy quantity of greater interest, namely the interaction energy, is then obtained as the difference between the total and self- energies of the system:
\[
U^{\rm int}(\lambda_N,\sigma_f) = U(\lambda_N,\sigma_f) - U^{\rm self}(\sigma_{f}).
\]

\subsection*{2. Electrostatic force}
Likewise in Lindgren et al. \cite{LindgrenManyBody2018}, the electrostatic force acting on each sphere follows the gradient of the (discrete) energy obtained in Appendix \ref{sect:elec_eng}, with respect to the coordinates of the centers. 
We will report here only the differences appearing by the presence of the ionic solvent and refer to such work \cite{LindgrenManyBody2018} for the general strategy of solving the forces based on the solution of an adjoint linear system. 

Indeed, it only remains to characterize the partial derivatives of $ A $ and $\bm f $ in this new setting.
According to \eqref{eq:Aii}, since $ n_i $ and $ L_i $ are independent on the particle coordinates, we have
\[
\partial_{x_k^\alpha} [A_{ii}]_{\ell \ell'}^{mm'} =  \partial_{x_k^\alpha} \left( 1 - \left[n_i\right]_\ell \left[L_i\right]_\ell \right) = 0,\qquad j=i,
\]
and according to \eqref{eq:Aij}, we have
\[
\partial_{x_k^\alpha} [A_{ij}]_{\ell \ell'}^{mm'} = \sum_{n=1}^{N} [A_{ij}^n ]_{\ell \ell'}^{m m'} \, 
\left( \partial_{x_k^\alpha} k_{\ell^\prime} (\kappa r_n^{ij}) \mathcal Y_{\ell^\prime m^\prime} (s_n^{ij}) + k_{\ell^\prime} (\kappa r_n^{ij}) \partial_{x_k^\alpha} \mathcal Y_{\ell^\prime m^\prime} (s_n^{ij}) \right),
\qquad j\neq i,
\]
where
\[
[A_{ij}^n ]_{\ell \ell'}^{m m'} = - \Ylm (s_n) w_n [n_j]_{\ell^\prime} [L_j]_{\ell^\prime} \dfrac{1}{k_{\ell^\prime} (\kappa r_j)}.
\]
The derivative of $k_{\ell^\prime} (\kappa r_n^{ij})$ should be computed as follows
\begin{align*}
	\partial_{x_k^\alpha} k_{\ell^\prime} (\kappa r_n^{ij}) & = \kappa k^\prime_{\ell^\prime} (\kappa r_n^{ij}) \partial_{x_k^\alpha} r_n^{ij}
	= \kappa k^\prime_{\ell^\prime} (\kappa r_n^{ij}) \dfrac{(v_n^{ij})_k}{r_n^{ij}} f_{ijk},
\end{align*}
where $ v_n^{ij} \coloneqq x_i+r_is_n - x_j$ and 
\[
f_{ijk} \coloneqq \left \{
\begin{aligned}
&1 && \mbox{if } k=i\neq j, \\
&-1 &&  \mbox{if } k=j\neq i, \\
&0 && \mbox{if } k=i = j, \\
&0 && \mbox{if } k\neq i \mbox{ and } k\neq j.
\end{aligned}
\right .
\]
In addition, we can compute the derivatives of $\bm f$ as follows
\[
\partial_{x_k^\alpha} 	[f_i]_\ell^m
=
\sum_{\substack{j\in \IntM \\ j\neq i}} 
\sum_{n=1}^{N_{\rm leb}} [f_j^n]_\ell^m \,\kappa \, k_0^\prime (\kappa r_n^{ij}) \dfrac{(v_n^{ij})_\alpha}{r_n^{ij}} f_{ijk},
\]
with
\[
[f_j^n]_\ell^m = \dfrac{1}{\varepsilon_\text{m}} \Ylm (s_n) \, w_n \, [n_j]_0 \dfrac{1}{k_0 (\kappa r_j)} \Yzz [\sigma_{f,i}]_0^0.
\]

\bibliography{Dielectric_PB_arXiv}

\end{document}